\documentclass[twocolumn,amsmath,aps,fleqn]{revtex4}
\usepackage{graphicx,amssymb}
\begin{document}
\newcommand{\be}{\begin{equation}}
\newcommand{\ee}{\end{equation}}
\newcommand{\bq}{\begin{eqnarray}}
\newcommand{\eq}{\end{eqnarray}}
\newcommand{\bsq}{\begin{subequations}}
\newcommand{\esq}{\end{subequations}}
\newcommand{\bc}{\begin{center}}
\newcommand{\ec}{\end{center}}
\newcommand {\R}{{\mathcal R}}
\newcommand{\al}{\alpha}
\newcommand\lsim{\mathrel{\rlap{\lower4pt\hbox{\hskip1pt$\sim$}}
    \raise1pt\hbox{$<$}}}
\newcommand\gsim{\mathrel{\rlap{\lower4pt\hbox{\hskip1pt$\sim$}}
    \raise1pt\hbox{$>$}}}

\title{Inner Structure of Black Holes in Eddington-inspired Born-Infeld gravity: the role of mass inflation}

\author{P.P. Avelino}
\email[Electronic address: ]{pedro.avelino@astro.up.pt}
\affiliation{Instituto de Astrof\'{\i}sica e Ci\^encias do Espa{\c c}o, Universidade do Porto, CAUP, Rua das Estrelas, PT4150-762 Porto, Portugal}
\affiliation{Centro de Astrof\'{\i}sica da Universidade do Porto, Rua das Estrelas, PT4150-762 Porto, Portugal}
\affiliation{Departamento de F\'{\i}sica e Astronomia, Faculdade de Ci\^encias, Universidade do Porto, Rua do Campo Alegre 687, PT4169-007 Porto, Portugal}

\date{\today}
\begin{abstract}

We investigate the interior dynamics of accreting black holes in Eddington-inspired Born-Infeld gravity using the homogeneous approximation and taking charge as a surrogate for angular momentum, showing that accretion can have an enormous impact on their inner structure. We find that, unlike in general relativity, there is a minimum accretion rate bellow which the mass inflation instability, which drives the centre-of-mass streaming density to exponentially high values in an extremely short interval of time, does not occur. We further show that, above this threshold, mass inflation takes place inside black holes very much in the same way as in general relativity, but is brought to a halt at a maximum energy density which is, in general, much smaller than the fundamental energy density of the theory. We conjecture that some of these results may be a common feature of modified gravity theories in which significant deviations from general relativity manifest themselves at very high densities.

\end{abstract}
\maketitle

\section{\label{intr}Introduction}

Eddington-inspired Born-Infeld (EiBI) gravity has been proposed as a solution to some of the astrophysical and cosmological singularities which occur in general relativity \cite{Banados:2010ix} (see also \cite{Vollick:2003qp,Vollick:2005gc,Vollick:2006qd}). Although in vacuum EiBI gravity is completely equivalent to Einstein's general relativity, it generates interesting new features in the presence of matter, specially for large densities or small lengthscales. These have been tightly constrained using astrophysical and cosmological observations \cite{Pani:2011mg,Casanellas:2011kf,Avelino:2012ge,Sham:2012qi,Harko:2013wka,Sotani:2014goa,Sotani:2014xoa} (see also \cite{Avelino:2012qe,EscamillaRivera:2012vz,Pani:2012qd,Sham:2013sya,Yang:2013hsa}  for potential shortcomings of EiBI gravity and \cite{Kim:2013nna,Avelino:2012ue} for possible solutions to some of them). Nevertheless, this theory may play a crucial dynamical role at the very high energies attained in the early universe \cite{Avelino:2012ue,Scargill:2012kg,Cho:2013pea,Bouhmadi-Lopez:2014jfa,Harko:2014nya,Cho:2014jta,Cho:2015yua} or inside black holes \cite{Sotani:2014lua,Jana:2015cha,Olmo:2013gqa,Harko:2013aya,Wei:2014dka,Shaikh:2015oha,Bazeia:2015uia}.

The inner structure of charged black holes in EiBI gravity has been investigated in \cite{Sotani:2014lua,Jana:2015cha} in the absence of accretion (see also \cite{Olmo:2013gqa,Harko:2013aya,Wei:2014dka,Shaikh:2015oha,Bazeia:2015uia}). However, realistic black holes are not isolated and are expected to rotate rapidly \cite{McClintock:2006xd,Brenneman:2006hw,Miller:2009cw,McClintock:2011zq,Brenneman:2011wz,Gou:2011nq}. Therefore, the computation of their internal structure needs to take into account accretion and angular momentum. In particular, an exponential growth of the Misner-Sharp mass, known as mass inflation, has been shown arise as a consequence of the relativistic counter-streaming between ingoing and outgoing streams inside charged Reissner-Nordstr$\ddot{\rm o}$m black holes and rotating Kerr black holes both in the context of general relativity \cite{Poisson:1989zz,Hod:1998gy,Ori:1991zz,Hansen:2005am,Hamilton:2008zz} and a number of modified gravity theories \cite{Avelino:2009vv,Avelino:2011ee,Hansen:2014rua,Avelino:2014aea,Hansen:2015dxa}. 

In this paper we shall investigate the inner structure of accreting spherically symmetric charged black holes, using charge as a surrogate for angular momentum. This is expected to be an excellent approximation given that the interior structure of a charged black hole resembles that of a rotating black hole, with the negative pressure associated to the electric field generating a gravitational repulsion analogous to that produced by the centrifugal force in a rotating black hole. 

This paper is organized as follows. We start by introducing EiBI gravity in Sec \ref{sec2}. Then, in Sec. \ref{EiBinoacc}, we consider spherically symmetric EiBI black hole solutions in the absence of accretion. The more general case of accreting EiBI black holes is introduced in Sec. \ref{EiBiacc} in the context of the homogeneous approximation. In Sec. \ref{disc} we present our results and discuss the role of mass inflation in EiBI black holes. We then conclude in Sec. \ref{conc}.

Throughout this paper we shall use fundamental units with $c=G=1$ and a metric signature $(-,+,+,+)$. The Einstein summation convention will be used when a greek index, taking the values $0,...,3$, appears twice in a single term (the exception will be the greek indices $\theta$ and $\phi$ which will denote the polar and azimuthal angles, respectively). 

\section{EiBI theory \label{sec2}}

The action for the EiBI theory of gravity is given by
\be
S=\frac{2}{\kappa}\int d^{4}x\left[\sqrt{\left|g_{\mu\nu}+\kappa R_{\mu\nu}(\Gamma)\right|}-\lambda\sqrt{|g|}\right]+S_M\,,\label{eq:EddingtonBornInfeld Action}
\ee
where $g_{\mu\nu}$ are the components of the metric, $g$ is the determinant of $g_{\mu\nu}$, $R_{\mu\nu}(\Gamma)$ is the symmetric Ricci tensor build from the connection $\Gamma$, and $S_M$ is the action associated with the matter fields. In the Palatini formulation the connection and the metric are treated as independent fields. 

Varying the action with respect to the connection yields the following equation of motion
\be
q_{\mu\nu}=g_{\mu\nu}+\kappa R_{\mu\nu}\,,\label{eq:ConnectionEquationOfMotion}
\ee
where $q_{\mu\nu}$ are the components of an auxiliary metric which is related to the connection by 
\be
\Gamma^{\gamma}_{\mu\nu} = {1 \over 2} q^{\gamma\zeta}(q_{\zeta\mu,\nu} + q_{\zeta\nu,\mu}- q_{\mu\nu,\zeta})\,, \label{connection}
\ee
and a comma represents a partial derivative. Varying the action with respect to the metric one obtains the other equation of motion
\be
\sqrt{|q|}q^{\mu\nu}=\lambda\sqrt{|g|}g^{\mu\nu}-{\bar \kappa}\sqrt{|g|}T^{\mu\nu}\,,\label{eq:MetricEquationOfMotion}
\ee
where $q^{\mu\nu}$ is the inverse of $q_{\mu\nu}$ and ${\bar \kappa}=8\pi \kappa$. Without loss of generality we set $\lambda=1$, since this term can be incorporated in the definition of the energy-momentum tensor. Combining Eqs. (\ref{eq:ConnectionEquationOfMotion}) and (\ref{eq:MetricEquationOfMotion}) one obtains
\be
{{{\mathcal R}}^\mu}_\nu \equiv q^{\mu \zeta} R_{\zeta \nu} =8\pi{\Theta^{\mu}}_{\nu}\,,\label{eq:EquationOfMotionComb}
\ee
where
\be
{\Theta^{\mu}}_{\nu}=\frac{1}{{\bar \kappa}}\left(1-\sqrt{{\left|\frac{g}{q}\right|}}\right){\delta^\mu}_\nu+\sqrt{{\left|\frac{g}{q}\right|}}{T^{\mu}}_{\nu} \,.\label{eq:Theta}
\ee
Using the above equations it is also possible to show that
\be
{{\mathcal G}^\mu}_\nu \equiv {{{\mathcal R}}^\mu}_\nu -\frac12 {\mathcal R} {\delta^\mu}_\nu  =8\pi {{\mathcal T}^{\mu}}_{\nu}\,,\label{eq:EquationOfMotionComb1}\,,
\ee
where 
\bq
{{\mathcal T}^{\mu}}_{\nu} &\equiv& {\Theta^{\mu}}_{\nu}-\frac12\Theta {\delta^\mu}_\nu\,,\\
\Theta &\equiv&  {\Theta^{\mu}}_{\mu}\,.
\eq

\section{\label{EiBinoacc} EiBI black holes: no accretion}

Let us start by describing the inner structure of charged EiBI black holes in the absence of accretion. The spherically symmetric physical ($g$) and auxiliary ($q$) line elements may be written as
\bq
ds_g^2&=&g_{tt}(r) dt^2+g_{rr}(r) dr^2+r^2(d\theta ^2+\sin ^2\theta d\phi ^2)\,, \label{le1}\\
ds_q^2&=&A(r) dt^2+B(r) dr^2+H^2(r)(d\theta ^2+\sin ^2\theta d\phi ^2)\,, \label{le2}
\eq
where $g_{tt}$, $g_{rr}$, $A\equiv q_{tt}$, $B \equiv q_{rr}$, and $H^2 \equiv g_{\theta \theta}$ are all functions of $r$ alone. 

The non-zero components of the energy-momentum tensor of the electric field corresponding to a constant charge $Q$ are given by
\bq
{^eT^r}_r&=&-\rho_e\,,  {^eT^t}_t=w_{e\parallel} \rho_e \,, \\
{^eT^\theta}_\theta&=&{^eT^\phi}_\phi=w_{e\perp} \rho_e\,,
\eq
with
\be
w_{e\parallel}=-1\,, \qquad w_{e\perp}=1\,,
\ee
and
\be
\rho_e=\frac{Q^2}{8\pi r^4}\,. 
\ee

\begin{figure}
\includegraphics[width=3.4in]{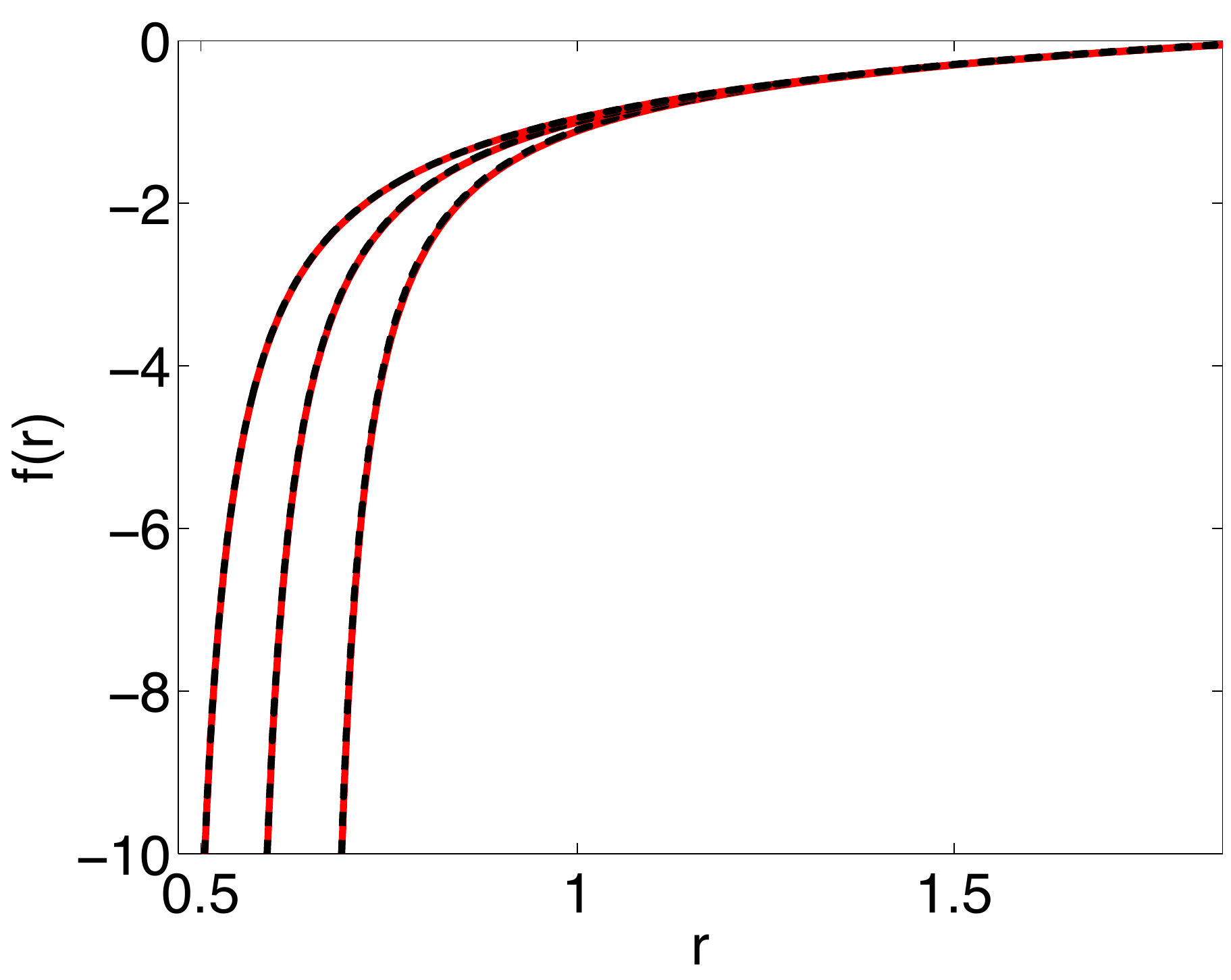}
\includegraphics[width=3.4in]{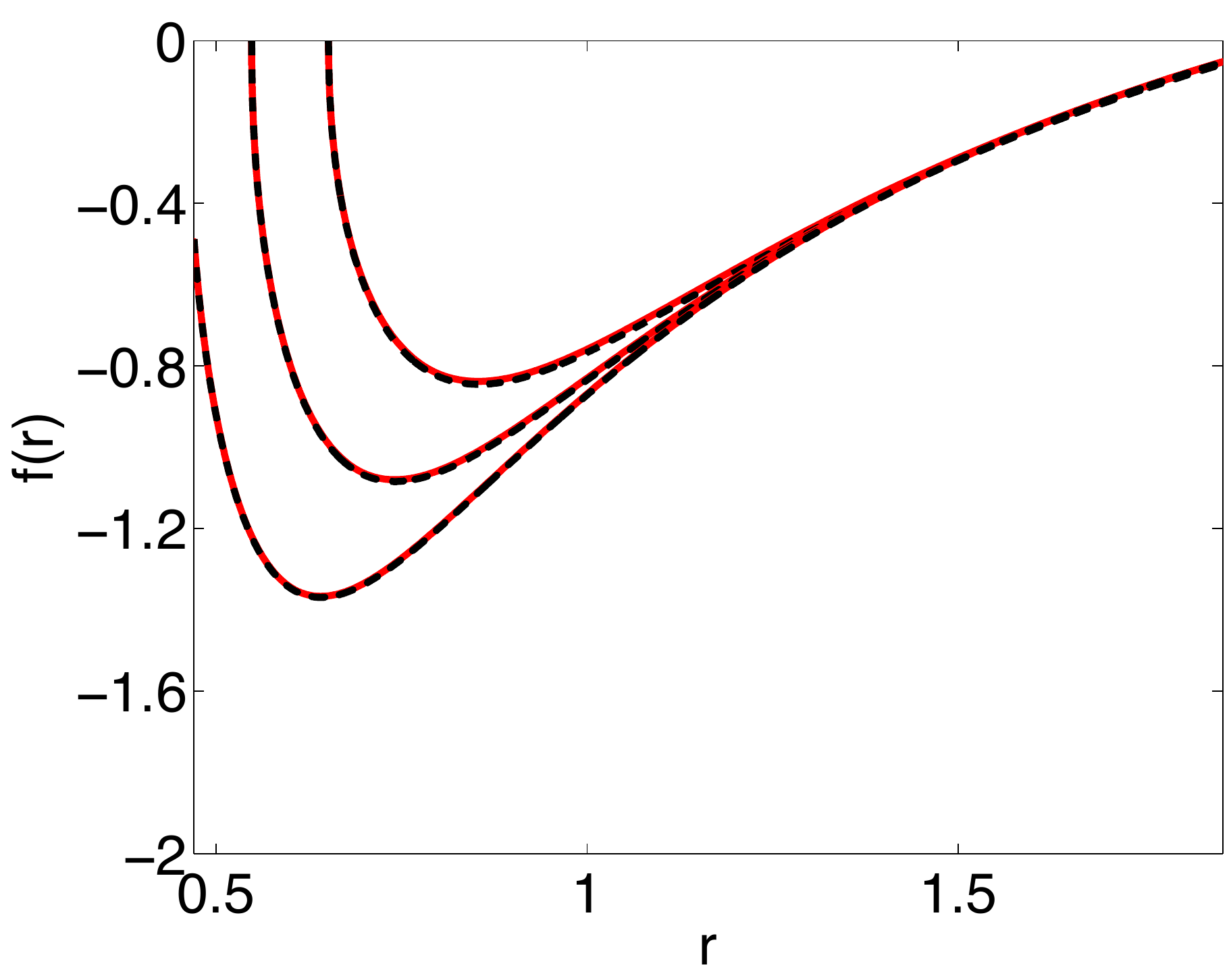}
\caption{Comparison between the function $f(r)$ obtained either analytically from Eq. (\ref{feq1}) (black dashed line) or using our numerical code (solid red line) for $\kappa=0.5$, $1$ and $2$ (from left to right, respectively), assuming that $M=1$ and $Q=0.3$. Notice the good agreement between the analytical and the numerical results.}
\label{Fig1}
\end{figure}

If the density is high enough, then the internal structure of EiBI black holes may be very different from their general relativity counterpart. In this case the following analytical solutions have been found \cite{Banados:2010ix,Sotani:2014lua,Jana:2015cha}
\bq
A&=& -\frac{r^4}{r^4 + \kappa Q^2} f\,, \\ 
B&=& \frac{1}{f}\,
\eq
with
\be
f = -\frac{r\sqrt{r^4 + \kappa Q^2}}{r^4 - \kappa Q^2}\int \frac{(- r^2 + Q^2)(r^4 - \kappa Q^2)}{r^4\sqrt{r^4 + \kappa Q^2}}dr \label{feq}\,.
\ee
Evaluating the integral in Eq. (\ref{feq}) and imposing that the solution approaches the standard Reissner-Nordstr$\ddot{\rm o}$m solution in the $r \to \infty$ limit one obtains that
\bq
f &=& -\frac{r\sqrt{r^4 + \kappa Q^2}}{r^4 - \kappa Q^2}  \left[   \left(3 \beta r^3  \sqrt{\kappa Q^2+r^4}\right)^{-1}\right. \nonumber\\
&\times& \left[\beta(Q^2-3r^2)(\kappa Q^2+r^4) - 4 i Q^2 r^3 \sqrt{\frac{r^4}{\kappa Q^2}+1} \right.\nonumber \\
&\times& \left.\left. \left(F\left[i{\rm arcsinh}\left(\beta r\right),-1\right]-\frac{\Gamma^2(1/4)}{4{\sqrt {i\pi }}}\right)\right]+2M\right]\,, \label{feq1}
\eq
where
\be
\beta \equiv \sqrt{\frac{i}{\sqrt \kappa Q}}\,,
\ee
$F$ represents the incomplete elliptic integral of the first kind, and, in this solution, $\Gamma$ denotes the gamma function.

Fig. \ref{Fig1} shows a comparison between the function $f(r)$ obtained either analytically from Eq. (\ref{feq1}) (black dashed line) or using our numerical code (solid red line) --- the numerical results will be discussed in detail in Sec. \ref{disc}. The results shown in Fig. \ref{Fig1} were obtained for $\kappa=0.5$, $1$ and $2$ (from left to right, respectively), assuming that $M=1$ and $Q=0.3$. Notice the good agreement between the analytical and the numerical results.

Fig. \ref{Fig1} also shows that, for $r \gg r_-$, the solution approaches the standard Reissner-Nordstr$\ddot{\rm o}$m solution characterized by the line element given in Eq. (\ref{le2}) with
\bq
A&=&g_{tt}=-\left(1-\frac{2M}{r}+\frac{Q^2}{r^2}\right)\,, \\ 
B&=&g_{rr}=-\frac{1}{A}\, \\ 
H&=&r\,,
\eq
This solution describes the space-time geometry in and around spherically symmetric black holes of mass $M$ and charge $Q$, and has inner ($r_-$) and outer ($r_-$) horizons located at
\be
r_{\pm}=\left(M \pm {\sqrt{M^2-Q^2}}\right)\,.
\ee
This result will be useful in Sec. \ref{disc} when discussing the initial conditions for the more general problem of the computation of the internal structure of charged spherically symmetric accreting black holes. 

\section{\label{EiBiacc} Accreting EiBI black holes}

The interior structure of black holes can be dramatically affected by accretion.  Here, we shall use the homogeneous approximation in the computation of the black hole's interior structure in the presence of accreting fluids, thus assuming that all relevant quantities can be written as a function of a radial (timelike) coordinate alone, as in Eqs. (\ref{le1}) and (\ref{le2}). This approximation, which is equivalent to the assumption of equal ingoing and outgoing streams, not only simplifies the mathematics but has also been shown to provide an accurate description of some of the most important aspects of mass inflation.

We shall now consider the most general fluid energy-momentum tensor consistent with spherical symmetry for which the homogeneous approximation remains valid. Its non-zero components are given by
\bq
{^fT^r}_r&=&-\rho_f\,,  {^fT^t}_t=w_{f\parallel} \rho_f \,, \\
{^fT^\theta}_\theta&=&{^fT^\phi}_\phi=w_{f\perp} \rho_f\,,
\eq
where $w_{f\parallel}=p_{f\parallel}/\rho_f$ and  $w_{f\perp}=p_{f\perp}/\rho_f$. Energy-momentum conservation of the fluid component implies that
\be
\frac{\rho_f'}{\rho_f}=-\frac{1+w_{f\parallel}}{2} \frac{g_{tt}'}{g_{tt}}-\frac{2(1+w_{f\perp})}{r}\label{rhoeq}\,,
\ee
where a prime represents a derivative with respect to the time-like coordinate $r$. Integrating Eq. (\ref{rhoeq}) with respect to $r$ one obtains
\bq
\rho_f&=& - {^fT^{r}}_{r}=\nonumber\\
&=&\rho_{fi}\left(\frac{g_{tti}}{g_{tt}}\right)^{(1+w_{f\parallel})/2} \left(\frac{r_i}{r}\right)^{2(1+w_{f\perp})}\,.
\eq
The total energy-momentum tensor
\be
{T^\mu}_\nu={^fT^\mu}_\nu+{^eT^\mu}_\nu\,,
\ee
is the sum of the fluid and electromagnetic parts, which are both assumed to be separately conserved. Consistently with the above notation, one may also write
\bq
{T^r}_r&=&-\rho\,,  {T^t}_t=w_\parallel \rho \,, \\
{T^\theta}_\theta&=&{T^\phi}_\phi=w_\perp \rho\,,
\eq
with $w_\parallel=p_\parallel/\rho$, $w_\perp=p_\perp/\rho$, $\rho=\rho_e+\rho_f$ and $p=p_e+p_f$.

The following relations between the components of the physical and auxiliary metrics can be computed using Eqs. (\ref{eq:ConnectionEquationOfMotion}) and (\ref{eq:MetricEquationOfMotion})
\bq
A&=&g_{tt} \frac{(1+{\bar \kappa} \rho)^{1/2} (1-{\bar \kappa} w_\perp \rho)}{(1-{\bar \kappa} w_\parallel \rho)^{1/2}}\,, \label{Avsgtt}\\ 
B&=&g_{rr} \frac{(1-{\bar \kappa} w_\parallel \rho)^{1/2} (1-{\bar \kappa} w_\perp \rho)}{(1+{\bar \kappa} \rho)^{1/2}}\,,\label{Bvsgrr}\\
H&=&r(1+{\bar \kappa}\rho)^{1/4}(1-{\bar \kappa} w_\parallel\rho)^{1/4}\,, \label{Hvsr}
\eq
and they imply that
\be
\sqrt{{\left|\frac{g}{q}\right|}}=\left(1+{\bar \kappa} \rho\right)^{-1/2} \left(1-{\bar \kappa} w_\parallel \rho\right)^{-1/2} \left(1-{\bar \kappa} w_\perp \rho\right)^{-1}\,.
\ee
Note that, in the absence of accretion, $\rho=\rho_e$, $p=p_e$, $w_\parallel=w_{e\parallel}$ and $w_\perp=w_{e\perp}$.

For ${\bar \kappa} \rho \ll 1$, and if the space-time gradients of the density field are small, the components of the physical ($g$) and auxiliary ($q$) metrics are approximately equal, and the non-zero components of ${{\mathcal T}^{\mu}}_{\nu}$ are given approximately by
\bq
{{\mathcal T}^t}_t&=& w_\parallel \rho = {T^t}_t\,, \\ 
{{\mathcal T}^r}_r&=& -\rho = {T^r}_r\,, \\
{{\mathcal T}^\theta}_\theta&=& w_\perp \rho = {T^\theta}_\theta\,, \\
{{\mathcal T}^\phi}_\phi&=& w_\perp \rho = {T^\phi}_\phi\,.
\eq
This is to be expected since in vacuum the EiBI theory is indistinguishable from general relativity.

\begin{figure}
\includegraphics[width=3.4in]{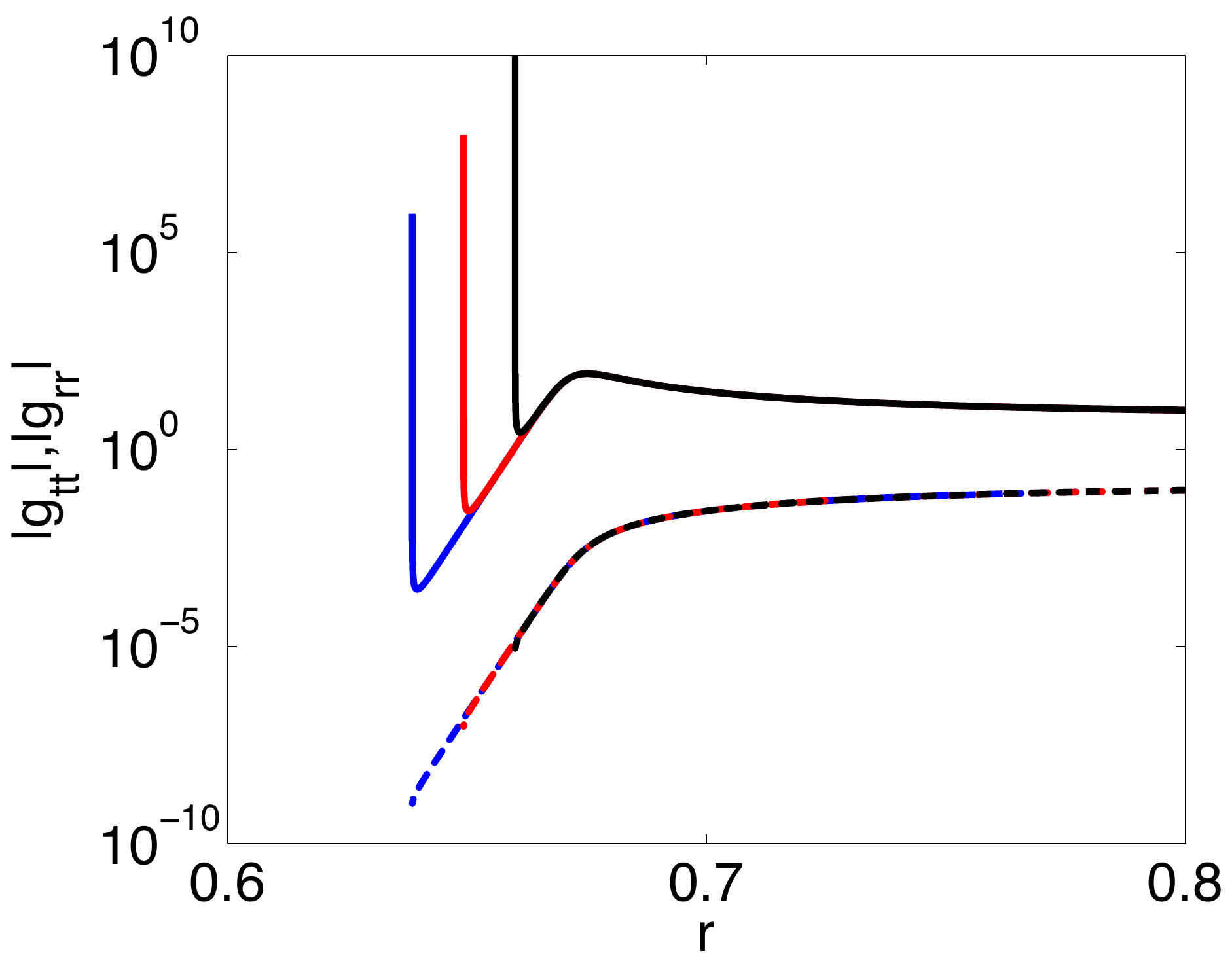}
\includegraphics[width=3.4in]{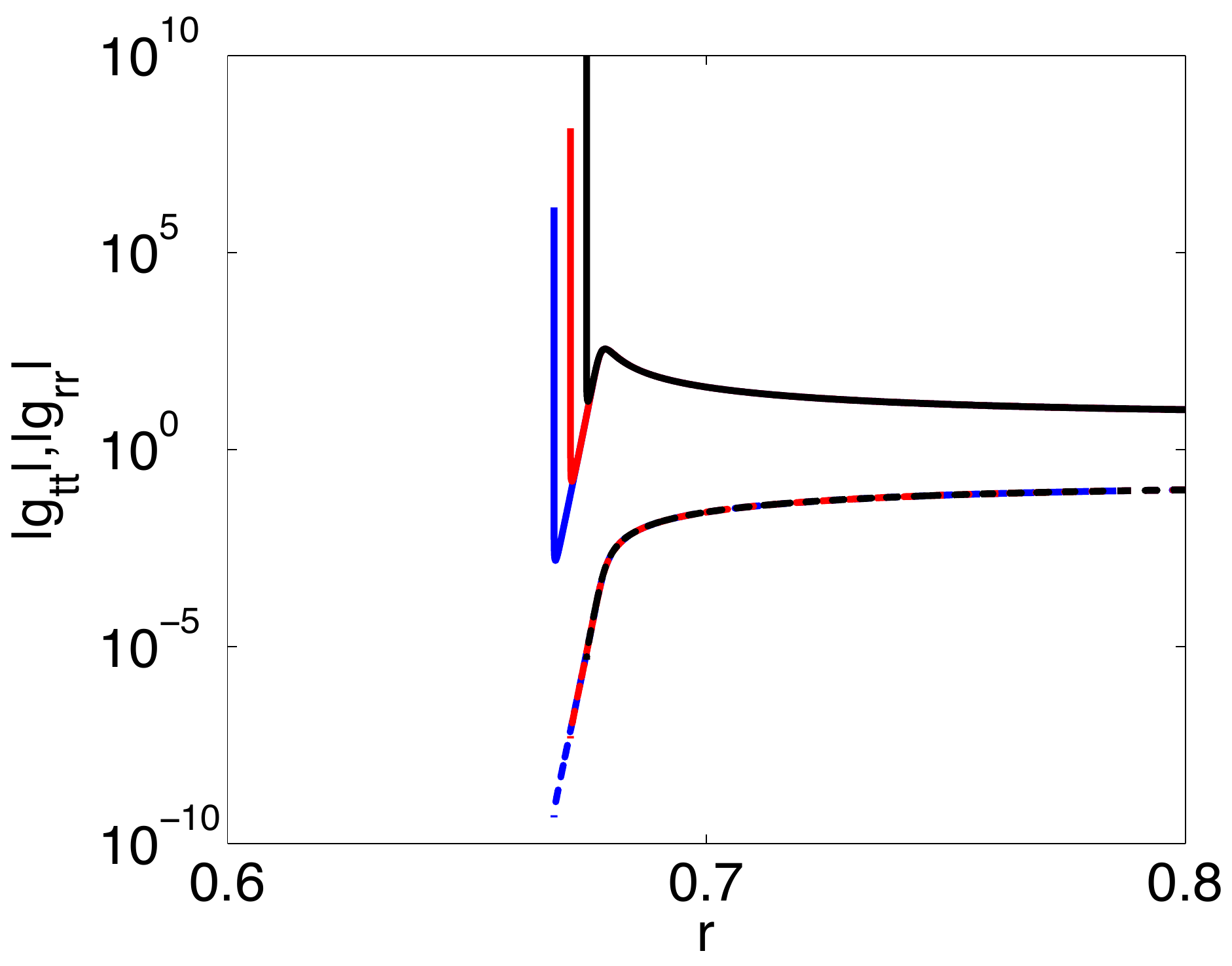}
\caption{The evolution of $|g_{rr}|$ and $|g_{tt}|$ with $r$ (solid and dashed lines, respectively) assuming that  $\rho_{fi}=10^{-3}$ and $\kappa=10^{-8}$, $10^{-6}$ and $10^{-4}$ (blue, red and black lines, from left to right, respectively). The top (bottom) panel assumes that $w_{f\perp}=1$ ($w_{f\perp}=0$). For $\kappa=10^{-8}$ and $\kappa=10^{-6}$ the behaviour of $|g_{rr}|$ for $r \lsim r_-$ indicates that mass inflation takes place inside the black hole  before it is brought to a halt at some value of $r<r_-$. For $\kappa=1 \times 10^{-4}$ no significant mass inflation takes place.}
\label{Fig2}
\end{figure}

\begin{figure}
\includegraphics[width=3.4in]{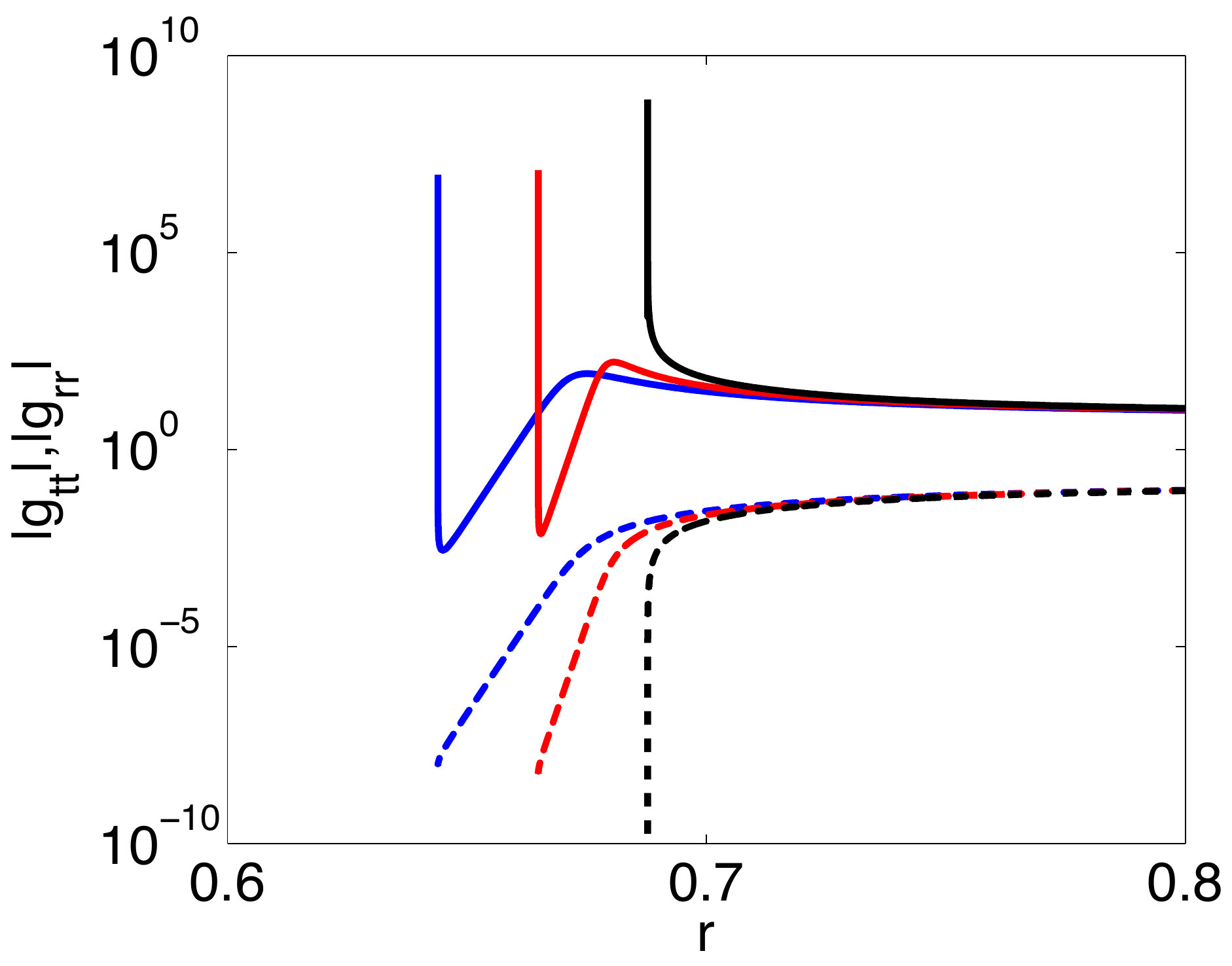}
\includegraphics[width=3.4in]{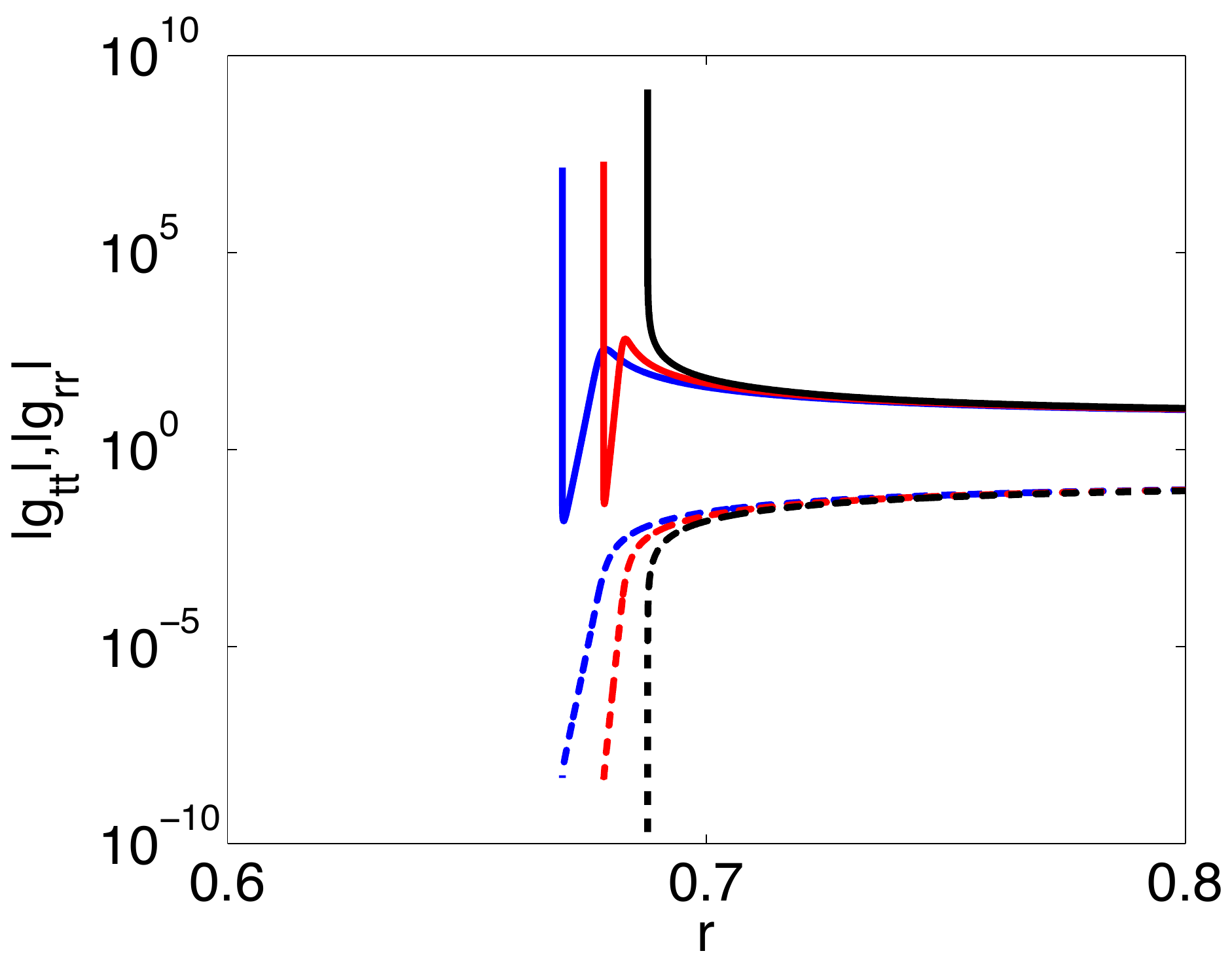}
\caption{The evolution of $|g_{rr}|$ and $|g_{tt}|$ with $r$ (solid and dashed lines, respectively) assuming that  $\kappa=10^{-7}$ and $\rho_{fi}= 10^{-7}$, $5 \times 10^{-4}$ and $10^{-3}$ (blue, red and black lines, from left to right, respectively). The top (bottom) panel assumes that $w_{f\perp}=1$ ($w_{f\perp}=0$). For $\rho_{fi}=5 \times 10^{-4}$ and $\rho_{fi}=10^{-3}$ the behaviour of $|g_{rr}|$ for $r \lsim r_-$ indicates that mass inflation takes place inside the black hole before it is brought to a halt at some value of $r<r_-$. For  $\rho_{fi}= 10^{-7}$ mass inflation does not occur.}
\label{Fig3}
\end{figure}

\begin{figure}
\includegraphics[width=3.4in]{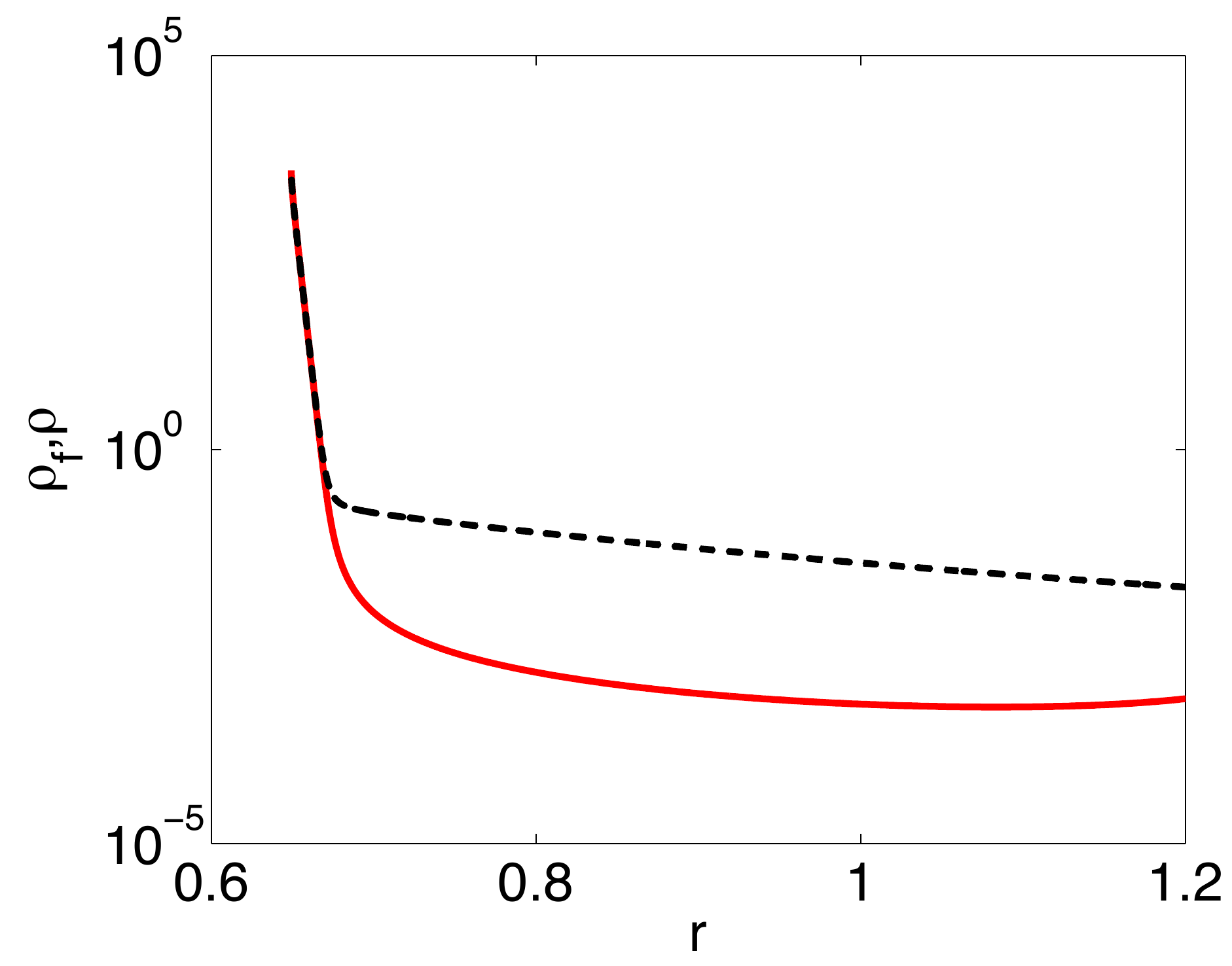}
\caption{The evolution of the total energy density $\rho$ (black dashed line) and of the fluid density $\rho_f$ with $r$ assuming $\rho_{fi}= 10^{-3}$, $\kappa=10^{-6}$ and $w_{f\perp}=1$. For large $r$ the fluid energy density $\rho_f$ is subdominant with respect to the electromagnetic energy density $\rho_e$, while for $r \lsim r_-$ it becomes the dominant component. Notice that mass inflation,  and the corresponding exponential growth of the energy density, is brought to a halt at a value of $\rho$ significantly smaller than the fundamental energy density of the theory ${\bar \kappa}^{-1}$.}
\label{Fig4}
\end{figure}

\begin{figure}
\includegraphics[width=3.4in]{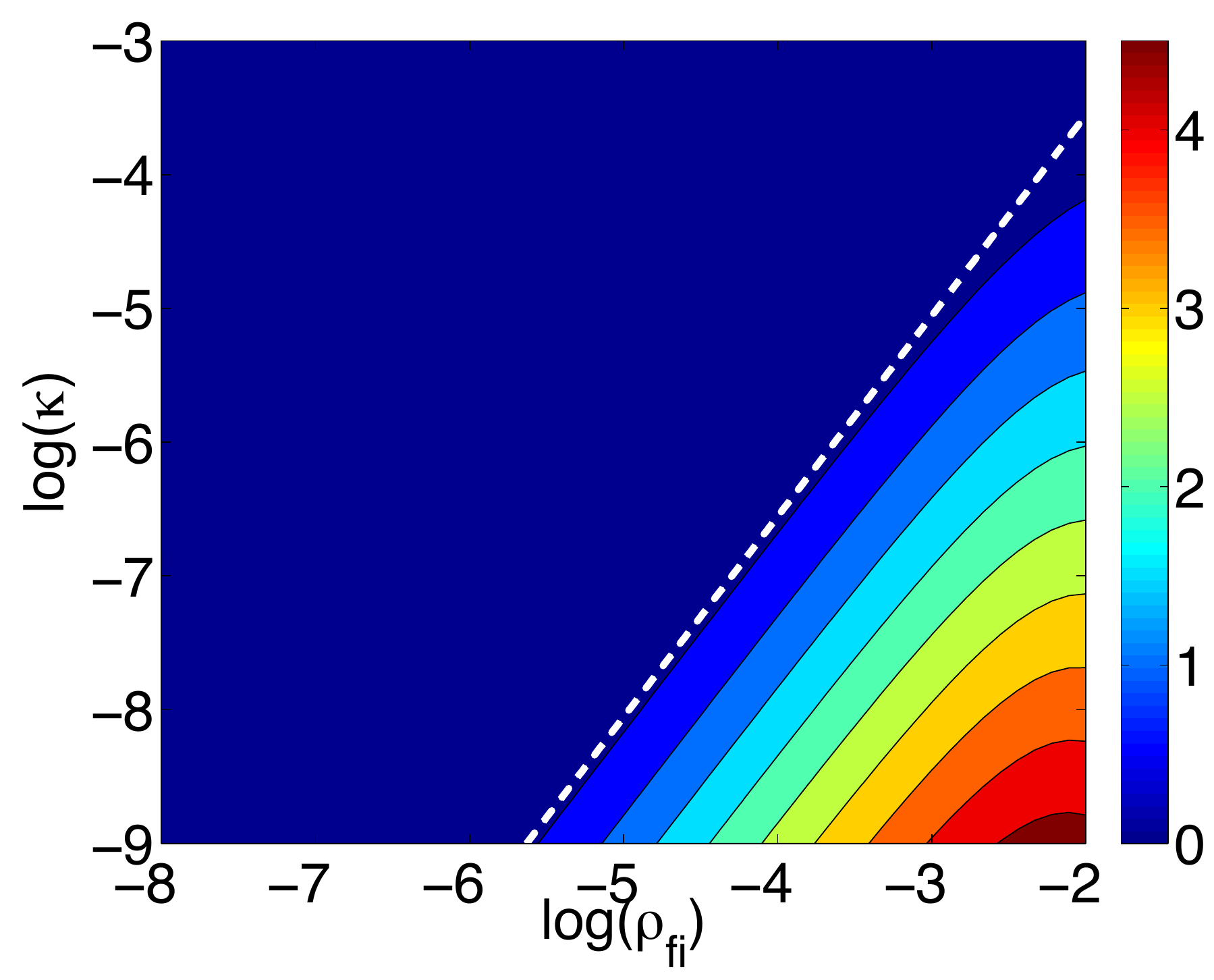}
\includegraphics[width=3.4in]{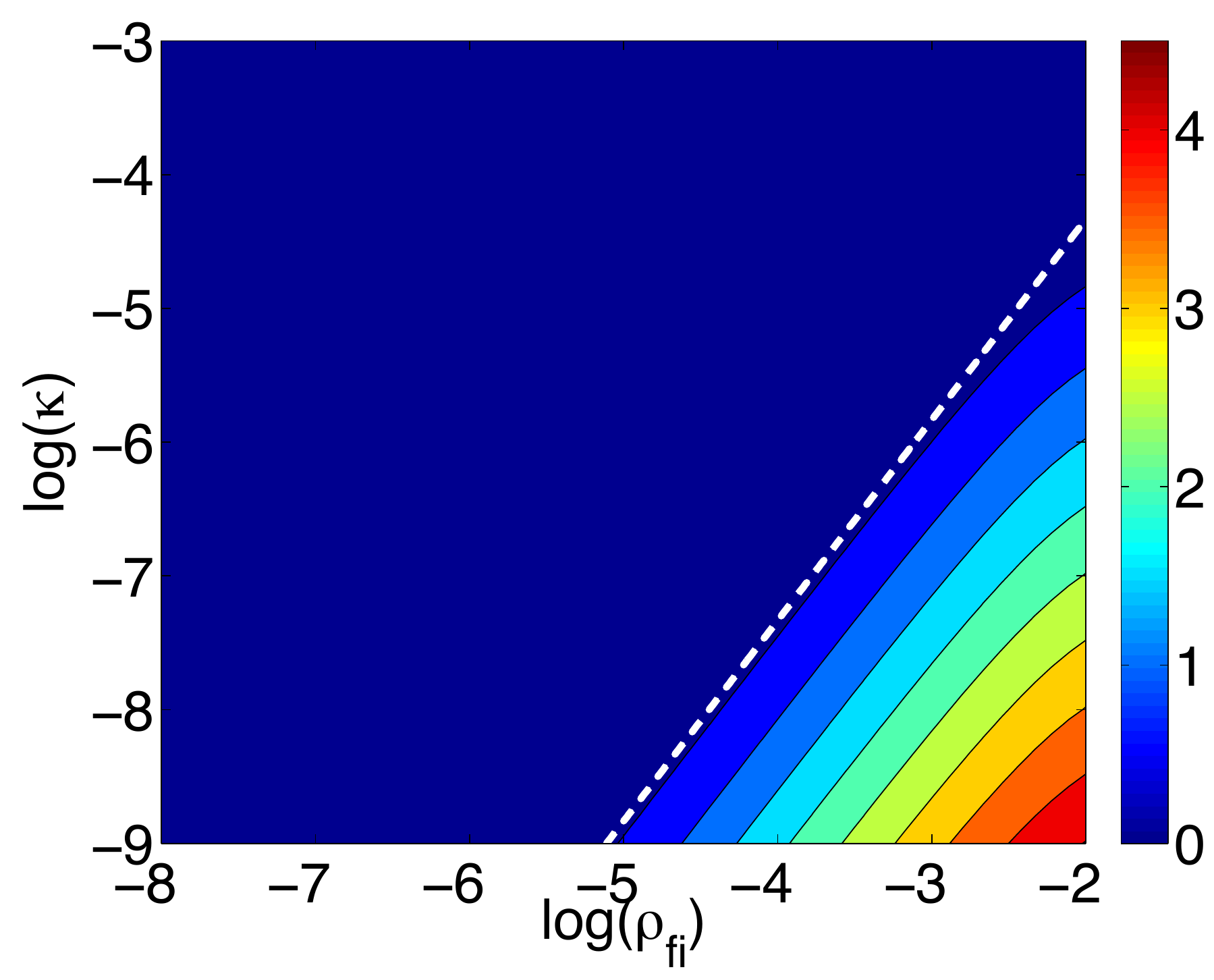}
\caption{The maximum value of the Misner-Sharp mass $M^{\rm max}_{\rm M-S}$ as a function of $\log(\rho_{fi})$ and $\log(\kappa)$, assuming that $w_{f\perp}=1$ (top panel) or $w_{f\perp}=0$ (lower panel). For small values of $\rho_{fi}$ the constant $M^{\rm max}_{\rm M-S}$ contours have a slope very close to $3/2$, indicating that  $M^{\rm max}_{\rm M-S}$ is a function of $\rho_{fi}^{3/2}/\kappa$. No significant mass inflation occurs to the left of the white dashed lines (whose slope is exactly $3/2$).}
\label{Fig5}
\end{figure}

\begin{figure}
\includegraphics[width=3.4in]{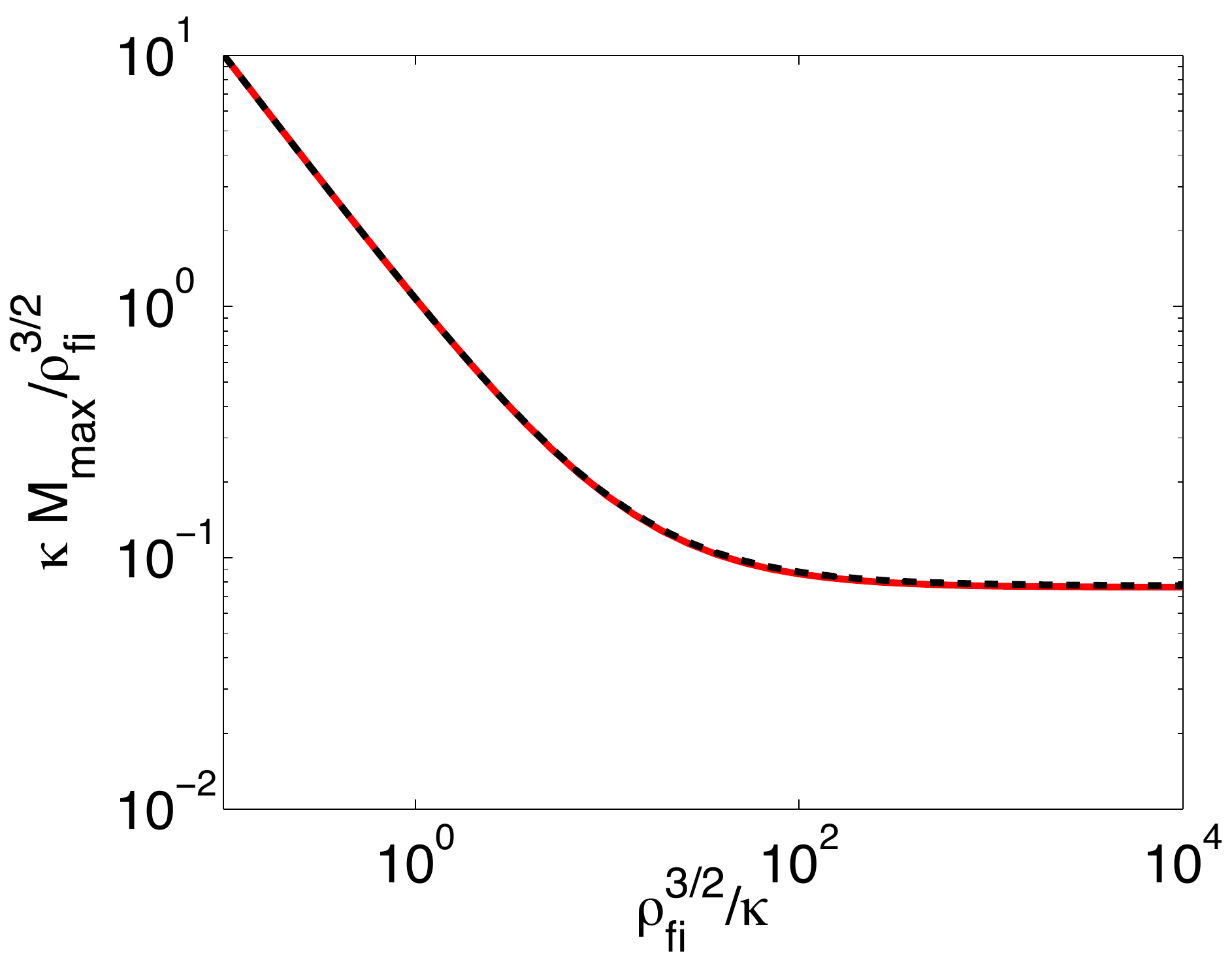}
\caption{The value of ${\kappa} M^{\rm max}_{\rm M-S}/\rho_{fi}^{3/2}$ as a function of $\rho_{fi}^{3/2}/\kappa$ assuming $\rho_{fi}=10^{-4}$ (red solid line) or $\kappa=10^{-11}$ (black dashed line), and $w_{f\perp}=0$. Fig. \ref{Fig6} shows that $M^{\rm max}_{\rm M-S}$ approaches unity at small values of $\rho_{fi}^{3/2}/\kappa$, becoming proportional to $\rho_{fi}^{3/2}/\kappa$ when $\rho_{fi}^{3/2}/\kappa$ is sufficiently large.}
\label{Fig6}
\end{figure}

\section{\label{disc} EiBI black holes: the role of mass inflation}

In the context of general relativity mass inflation has been shown to be generic due to the relativistic counter-streaming between ingoing and outgoing streams driving $w_{f\parallel}$ towards unity during mass inflation. For accretion of a massless scalar field $w_{f\perp}=w_{f\parallel}=1$ while for relativistic particles $w_{f\perp}=(1-w_{f\parallel})/2$ which becomes tiny in the mass inflation region. Hence, in the remainder of this paper we shall take $w_{f\parallel}=1$ and consider the fiducial cases with $w_{f\perp}=1$ or $w_{f\perp}=0$.

Without loss of generality, we shall choose units such that $M=1$. Furthermore, given that observational evidence suggest that real astrophysical black holes may have nearly extremal spins \cite{McClintock:2006xd,Brenneman:2006hw,Miller:2009cw,McClintock:2011zq,Brenneman:2011wz,Gou:2011nq}, we shall take $Q=0.95$, close to the extremal value of unity, in the remainder of the paper (note that, for this particular choice of $M$ and $Q$, $r_- \sim 0.69$ and $r_+ \sim 1.31$). 

The $tt$ components of Eqs. (\ref{eq:EquationOfMotionComb}) and (\ref{eq:EquationOfMotionComb1}) may be written, respectively, as 
\bq
A''&=&\frac{A'^2}{2A} +\frac{A'B'}{2B}-\frac{2A'H'}{H}-16 \pi  {\Theta^t}_t  A B  \label{Aeq}\,,\\
B'&=&-\frac{B^2}{HH'} +\frac{BH'}{H}+\frac{2BH''}{H'}-8\pi {{\mathcal T}^r}_r  \frac{B^2 H}{H'} \label{Beq}\,.
\eq
We solve numerically the coupled system of differential equations given by Eqs. (\ref{rhoeq}), (\ref{Aeq}) and (\ref{Beq}), with $g_{tt}$ given by Eq. (\ref{Avsgtt}), using a fourth order Runge-Kutta scheme with adaptive step size. The value of $H$ was  calculated, as a function of $r$ and $\rho$, directly from Eq. (\ref{Hvsr}). The equations for $H'$ and $H''$ are not explicitly provided given their very large size. Nevertheless, the value of $H'$ may be obtained, as a function of $r$, $\rho$, $A$ and $A'$, by differentiating Eq. (\ref{Hvsr}) with respect to $r$ and using Eq.  (\ref{rhoeq}).  The value of $H''$ may also be found,  as a function of $r$, $\rho$, $A$, $B$ and $A'$, by differentiating the equation for $H'$ with respect to $r$ and using Eqs. (\ref{rhoeq}), (\ref{Aeq}) and (\ref{Beq}).

We consider initial conditions with $r_i=0.95 r_-$ and assume that ${\bar \kappa} \ll 1$. In this limit, if the accretion rate is not unreasonably large, the inner structure of the black hole around $r=r_i$ is very close to that of an ordinary charged Reissner-Nordstr$\ddot{\rm o}$m black hole in general relativity. Consequently, our initial conditions satisfy
\bq
A(r_i)&=&-\left(1-\frac{2M}{r_i}+\frac{Q^2}{r_i^2}\right)\,,\\
B(r_i)&=&-\frac{1}{A(r_i)}\,,\\
H(r_i)&=&r_i\,,\\
\rho_{f}(r_i)&=&\rho_{fi}\,,
\eq
where $\rho_{fi}$ parameterizes the accretion rate.

Fig. \ref{Fig2} shows the evolution of $|g_{rr}|$ and $|g_{tt}|$ with $r$ (solid and dashed lines, respectively) computed using our numerical algorithm assuming $\rho_{fi}=10^{-3}$ and $\kappa=10^{-8}$, $10^{-6}$ and $10^{-4}$ (blue, red and black lines, from left to right, respectively). The results presented in the top (bottom) panel assume that $w_{f\perp}=1$ ($w_{f\perp}=0$).  For $\kappa=10^{-8}$ and $\kappa=10^{-6}$ the behaviour of $|g_{rr}|$ for $r \lsim r_-$ shows that there is an exponential growth of the Misner-Sharp mass
\bq
M_{\rm M-S}=\frac{r}{2}\left(1+\frac{Q^2}{r^2}-g_{rr}^{-1}\right)\,.
\eq
Consequently, in these cases mass inflation takes place inside the black hole before it is brought to a halt at some value of $r<r_-$. On the other hand, for $\kappa=1 \times 10^{-4}$, no significant mass inflation takes place.

Fig. \ref{Fig3} shows the evolution of $|g_{rr}|$ and $|g_{tt}|$ with $r$ (solid and dashed lines, respectively) computed using our numerical code assuming $\kappa=10^{-7}$ and $\rho_{fi}= 10^{-7}$, $5 \times 10^{-4}$ and $10^{-3}$ (blue, red and black lines, from left to right, respectively). As in Fig. \ref{Fig2}, the top (bottom) panel assumes that $w_{f\perp}=1$ ($w_{f\perp}=0$). For $\rho_{fi}=5 \times 10^{-4}$ and $\rho_{fi}=10^{-3}$ the behaviour of $|g_{rr}|$ for $r \lsim r_-$ indicates that mass inflation takes place inside the black hole  before it is brought to a halt at some value of $r<r_-$, while for  $\rho_{fi}= 10^{-7}$ mass inflation does not occur.

Fig. \ref{Fig4} illustrates the evolution of the total energy density $\rho$ (black dashed line) and of the fluid density $\rho_f$ with $r$ assuming $\rho_{fi}= 10^{-3}$, $\kappa=10^{-6}$ and $w_{f\perp}=1$. It shows that while for large values of $r$ the fluid energy density $\rho_f$ is subdominant with respect to the electromagnetic density $\rho_e$, for $r \lsim r_-$ it becomes the dominant component ($\rho \sim \rho_f$ during mass inflation). Mass inflation, and the corresponding exponential growth of the energy density,  is brought a halt at a value of the energy density significantly smaller than the fundamental energy density of the theory ${\bar \kappa}^{-1}$. We verified that this is always the case, unless the accretion rate, parameterized by $\rho_{fi}$, is extremely large.

Fig. \ref{Fig5} displays the maximum value of the Misner-Sharp mass $M^{\rm max}_{\rm M-S}$ as a function of $\log(\rho_{fi})$ and $\log(\kappa)$, assuming $w_{f\perp=1}$ (top panel) and $w_{f\perp}=0$ (bottom panel). For small values of $\rho_{fi}$ the constant $M^{\rm max}_{\rm M-S}$ contours have a slope very close to $3/2$, indicating that  $M^{\rm max}_{\rm M-S}$ is a function of $\rho_{fi}^{3/2}/\kappa$. In both the top and bottom panels, significant mass inflation occurs only for sufficiently high values of $\rho_{fi}^{3/2}/\kappa$ (or, equivalently, to the right of the white dashed lines whose slope is exactly $3/2$).

The value of ${\kappa} M^{\rm max}_{\rm M-S}/\rho_{fi}^{3/2}$ is plotted in Fig. \ref{Fig6} as a function of $\rho_{fi}^{3/2}/\kappa$ assuming $\rho_{fi}=10^{-4}$ (red solid line) or $\kappa=10^{-11}$ (black dashed line). Fig. \ref{Fig6} shows that $M^{\rm max}_{\rm M-S}$ approaches unity at small values of $\rho_{fi}^{3/2}/\kappa$, becoming proportional to $\rho_{fi}^{3/2}/\kappa$ when $\rho_{fi}^{3/2}/\kappa$ is sufficiently large. 

All the results presented in Sec. \ref{disc} of the paper assumed that $\kappa>0$. However, we verified that the results are only very weakly sensitive to the sign of $\kappa$, with the modifications to the figures associated to the sign change being barely noticeable.

\section{\label{conc} Conclusions}

In this paper we investigated the structure of accreting EiBI black holes using the homogeneous approximation and taking charge as a surrogate for angular momentum. We have shown that mass inflation inside EiBI black holes, if it occurs, is brought to an end at an energy density threshold significantly smaller than the fundamental energy density of the theory ${\bar \kappa}^{-1}$. We have further shown that the maximum value of the Misner-Sharp mass is roughly proportional to $\rho_{fi}^{3/2}/{\kappa}$ for small values of the accretion rate (parameterized by $\rho_{fi}$). This implies that mass inflation does not happen for low accretion rates, independently of how close the EiBI gravity is to general relativity (the $\kappa=0$ limit of EiBI gravity). 

Arbitrary deviations from general relativity in spherically symmetric spacetimes may be described in terms of modifications to the radial and transverse equation of state of an effective fluid component. Mass inflation requires the effective equation of state parameter in the parallel direction $w_\parallel$ to be close to unity. Consequently, mass inflation cannot proceed in the presence of modifications to general relativity which force $w_{\parallel}$ to deviate significantly from unity. Therefore, the abrupt end to the mass inflation regime observed in EiBI gravity is expected to be a common feature of modified gravity theories in which significant deviations from general relativity manifest themselves at very high densities.

\begin{acknowledgments}

The author thanks Andrew Hamilton and Carlos Herdeiro for enlightening discussions on the subject of mass inflation. This work was supported by Funda{\c c}\~ao para a Ci\^encia e a Tecnologia (FCT) through the Investigador FCT contract of reference IF/00863/2012 and POPH/FSE (EC) by FEDER funding through the program "Programa Operacional de Factores de Competitividade - COMPETE. Funding of this work was also provided by the FCT grant UID/FIS/04434/2013

\end{acknowledgments}


\bibliography{EiBI-MI}

\end{document}